\documentclass[preprint,showpacs,preprintnumbers,amsmath,amssymb]{revtex4}
\usepackage{epsfig}
\usepackage{graphicx}
\usepackage{dcolumn}
\usepackage{bm}

\def\beq{\begin{equation}}
\def\eeq{\end{equation}}
\def\eeqn{\end{equation}}
\newcommand\iden{\leavevmode\hbox{\small1\normalsize\kern-.33em1}}

\newcommand{\bea} {\begin{eqnarray}}
\newcommand{\eea} {\end{eqnarray}}

\let\jnfont=\rm
\def\NPB#1,{{\jnfont Nucl.\ Phys.\ B }{\bf #1},}
\def\PLB#1,{{\jnfont Phys.\ Lett.\ B }{\bf #1},}
\def\EPJC#1,{{\jnfont Eur.\ Phys.\ Jour.\ C }{\bf #1},}
\def\PRD#1,{{\jnfont Phys.\ Rev.\ D }{\bf #1},}
\def\PRL#1,{{\jnfont Phys.\ Rev.\ Lett.\ }{\bf #1},}
\def\MPLA#1,{{\jnfont Mod.\ Phys.\ Lett.\ A }{\bf #1},}
\def\JPG#1,{{\jnfont J.\ Phys.\ G }{\bf #1},}
\def\CTP#1,{{\jnfont Commun.\ Theor.\ Phys.\ }{\bf #1},}
\def\JHEP#1,{{\jnfont JHEP \ }{\bf #1},}
\def\NPPS#1,{{\jnfont Nucl.\ Phys.\ Proc.\ Suppl.\ }{\bf #1},}
\def\CPC#1,{{\jnfont Computl.\ Phys.\ Commun.\ }{\bf #1},}
\def\CPL#1,{{\jnfont Chin.\ Phys.\ Lett. }{\bf #1},}
\def\AJS#1,{{\jnfont Astrophys.\ J.\ Suppl. }{\bf #1},}
\def\PR#1,{{\jnfont Phys.\ Rept. }{\bf #1},}
\def\AP#1,{{\jnfont Astropart.\ Phys. }{\bf #1},}
\def\EPL#1,{{\jnfont Europhys.\ Lett. }{\bf #1},}
\def\FP#1,{{\jnfont Fortsch.\ Phys. }{\bf #1},}

\begin{document}

\title{\ \\[10mm] LHC diphoton Higgs signal and top quark forward-backward asymmetry
in quasi-inert Higgs doublet model}

\author{Lei Wang, Xiao-Fang Han}

\affiliation{ Department of Physics, Yantai University, Yantai
264005, China}


\begin{abstract}
In the quasi-inert Higgs doublet model, we study the LHC diphoton
rate for a standard model-like Higgs boson and the top quark
forward-backward asymmetry at Tevatron. Taking into account the
constraints from the vacuum stability, unitarity, electroweak
precision tests, flavor physics and the related experimental data of
top quark, we find that compared with the standard model prediction,
the diphoton rate of Higgs boson at LHC can be enhanced due to the
light charged Higgs contributions, while the measurement of the top
quark forward-backward asymmetry at Tevatron can be explained to
within $1\sigma$ due to the non-standard model neutral Higgs bosons
contributions. Finally, the correlations between the two observables
are discussed.

\end{abstract}



\maketitle

\section{Introduction}
LHC \cite{12010019-2-3,1202.1487,1202.1414} has recently reported
some hints for a 125 GeV Higgs boson decaying into two photons and
the diphoton rate should be larger than the standard model (SM) one.
Due to the absence of a true signal, the expected exclusion limit at
95\% confidence level is between 1.4 and 2.4 times the SM rate in
the mass range 110-150 GeV from the CMS collaboration
\cite{1202.1487} and between 1.6 and 1.7 times in the mass range
115-130 GeV from ATLAS collaboration \cite{1202.1414}. The LHC
diphoton signal has been studied in various extensions of the SM,
such as the minimal supersymmetric standard model (MSSM)
\cite{11125453-26272829}, the two Higgs doublet model
\cite{11125453-3234}, the inert Higgs doublet model \cite{12012644},
the next-to-MSSM \cite{11125453-3537}, the little Higgs models
\cite{11125453-3839}, the Higgs triplet models \cite{11125453-40}
and the models with extra dimension \cite{extrad}

The forward-backward asymmetry $A_{FB}^t$ in top quark pair
production has been measured by two experimental groups at the
Tevatron. The CDF measured the asymmetry in the $\ell+j$ channel and
obtained $A_{FB}^t(CDF)= 0.158\pm0.074$ \cite{11100014-1}, which is
nearly consistent with the D0 result $A_{FB}^t(D0) = 0.19\pm0.065$
\cite{11100014-2}. These results exceed the NLO SM prediction,
$A_{FB}^t(SM) = 0.058\pm 0.009$ \cite{11084005-4¨C7}. The CDF also
reported an anomaly large value of $A_{FB}^t$ for $m_{t\bar{t}}>
450$ GeV \cite{11081802-8}, which, however, is not confirmed by D0
collaboration \cite{11100014-2}. Therefore, we do not consider the
experimental data in this paper. To explain $A_{FB}^t$, various
attempts have been tried, such as via the $s$-channel exchange of an
axi-gluon \cite{afb-s-channel} or the $t$-channel exchange of $Z'$,
$W'$ and a scalar
\cite{afb-tu-channel,afbdefi,11074350,lrth,09113237}.

The quasi-inert Higgs doublet model (QIHDM) \cite{QIHDM} is proposed
by Qing-Hong Cao et al. with the motivation of explaining the excess
of $Wjj$ reported by CDF \cite{wjjcdf}. In addition to the SM
particle content, just one complex electroweak scalar doublet is
introduced with an approximate $Z_2$-symmetry being imposed. This
model predicts the SM-like Higgs boson $h$, the charged Higgs boson
$H^\pm$ as well as two neutral Higgs bosons $S$ and $A$. The three
new Higgs bosons can contribute to the diphoton rate of $h$ and the
top quark forward-backward asymmetry at Tevatron. Since D0
collaboration does not confirm the excess of $Wjj$ reported by CDF
\cite{wjjd0}, we will not discuss the experimental result of $Wjj$
in this paper.

 This work is organized as follows. In Sec. II, we briefly
review the quasi-inert Higgs doublet model. In Sec. III, we discuss
the relevant theoretical and experimental constraints. In Sec. IV,
we study diphoton rate for the SM-like Higgs boson at LHC. In Sec.
V, we study the top quark forward-backward asymmetry $A_{FB}^t$ at
Tevatron and the correlation between $A_{FB}^t$ and the LHC
diphoton Higgs signal. Finally, we give our conclusion in Sec. VI.

\section{quasi-inert Higgs doublet model}

The quasi-inert Higgs doublet model is a simple extension of SM by
including an additional scalar doublet
$\Phi_{2}=[H^{+},(S+iA)/\sqrt{2}]$, with the same hypercharge as the
SM doublet $\Phi_{1}=(\phi^{+},\phi^{0})$. To explicitly forbid the
mixing between the two Higgs doublets, we can impose a
$Z_2$-symmetry under which $\Phi_{2}$ and $\Phi_{1}$ are
respectively odd and even. The renormalizable scalar potential can
be written as \cite{QIHDM}
\begin{eqnarray}\label{potent}
V & = &
\mu_{1}^{2}\Phi_{1}^{\dagger}\Phi_{1}+\mu_{2}^{2}\Phi_{2}^{\dagger}
\Phi_{2} +\lambda_{1}\left(\Phi_{1}^{\dagger}\Phi_{1}\right)^{2}
+\lambda_{2}\left(\Phi_{2}^{\dagger}\Phi_{2}\right)^{2}
+\lambda_{3}\left(\Phi_{1}^{\dagger}\Phi_{1}\right)
 \left(\Phi_{2}^{\dagger}\Phi_{2}\right)\notag\\
 &  & +\lambda_{4}
 \left(\Phi_{1}^{\dagger}\Phi_{2}\right)
 \left(\Phi_{2}^{\dagger}\Phi_{1}\right)+\frac{1}{2}\lambda_{5}
 \left(\Phi_{1}^{\dagger}\Phi_{2}\right)^{2}+\frac{1}{2}\lambda_{5}^{*}
 \left(\Phi_{2}^{\dagger}\Phi_{1}\right)^{2},
\end{eqnarray}


All the above parameters are necessarily real except $\lambda_{5}$.
The SM $SU(2)\times U(1)_{Y}$ gauge symmetry is spontaneously broken
by the vacuum expectation value (VEV) of $\phi^{0}$ field,
$\left\langle \phi^{0}\right\rangle =v=246\,{\rm GeV}$. $\Phi_{2}$
field does not develop VEV from Eq. (\ref{potent}).

The quartic coupling $\lambda_i$ can be expressed by the physical
scalar masses and $\mu_2$ as follow:
\bea \lambda_1 &=& \frac{m_h^2}{2 v^2}\quad , \quad
\lambda_3 = \frac{2}{v^2}\left(m_{H^\pm}^2 - \mu_2^2\right), \nonumber\\
\lambda_4 &=& \frac{\left(m_S^2 + m_A^2 - 2 m_{H^\pm}^2\right)}{v^2}
\quad , \quad \lambda_5 = \frac{\left(m_S^2 - m_A^2\right)}{v^2}
\label{lambds} \eea The scalar potential in Eq. (\ref{potent})
contains the SM-like Higgs boson coupling with the charged Higgs,
\begin{equation} g_{hH^{\pm}H^{\mp}} = -i\lambda_3 v=-i\frac{2m^2_{H^\pm}}
{v}\left(\frac{m^2_{H^\pm}-\mu^2_2}{m^2_{H^\pm}}\right). \label{hcc}
\end{equation}

In order to explain the top quark forward-backward asymmetry at
Tevatron, $\Phi_{2}$ field has to couple to the first generation
quark. Therefore, we require the right-handed up quark is odd under
$Z_2$-symmetry, while all the other SM fields are even. Thus, the
Yukawa couplings of the quark fields can be written as \cite{QIHDM}
\begin{equation}\label{allyukawa}
\mathcal{L}_{\mathcal{Y}}=-y_{u,1}^{ik} \bar{Q}_L^i \tau_2
\Phi_{1}^* u^k_R - y_{d,1}^{ij}\bar{Q}_L^i\Phi_{1}d_R^{j} -
y^{i1}_{u,2}\bar{Q}_L^i \tau_2 \Phi_{2}^* u^1_R+h.c.\,,
\end{equation}
where $\tau_2=i\sigma_2$ ($\sigma_2$ is Pauli matrix) and
${Q}_L^i=\left(u_L^i,d_L^i\right)^T$. The generation indices $i$,
$j$ run from 1 to 3, and $k$ can only takes $2$ and $3$. The
coupling of the right-handed up quark to $\Phi_1$ is forbidden by
the $Z_2$-symmetry, which leads that the up quark remains massless
at tree-level. In order to generate the up quark mass, a soft
breakdown scalar potential of the $Z_2$-symmetry,
$\mu_{12}^2\Phi_{1}^{\dagger}\Phi_{2}+h.c.$, is necessary. Since the
up quark mass is about a few MeV, $\mu_{12}^2$ is relative small,
which can induce a negligible VEV of $\Phi_{2}$ field compared with
that of $\Phi_{1}$ field.

In the mass eigenstates, we can obtain the coupling of the
right-handed up quark to $\Phi_2$ field from Eq. (\ref{allyukawa}),
\bea {\cal L}_{u}=&-&\frac{S-iA}{\sqrt{2}}~(X_u)^{i1}~ \bar{u}_L^{i}
u_R^{1}+ H^- ~(V_{CKM}^\dag X_u)^{i1}~\bar{d}^{~i}_L u_R^{1}+h.c.,
 \label{yukhat}
\eea
where $(X_u)^{i1}=(U_{u,L}^\dag)^{ij} (y_{u,2})^{j1}$ with $U_{u,L}$
being the rotation matrix of the left-handed up-type quarks. We choose to make
the simplifying assumption, $(U_{u,L})^{ij}=\delta^{ij}$.

To explain the top quark forward-backward asymmetry and satisfy the
constraints from the flavor processes, we take \beq
(X_u)^{i1}=2y_1(V_{CKM})^{i3}.\eeq
The detailed analysis was given in \cite{11074350}.
Where $y_1=\frac{1}{2}(y_{u,2})^{i1}/(V_{CKM})^{i3}$ for $(U_{u,L})^{ij}=\delta^{ij}$.
From the Eq.
(\ref{yukhat}), we can get the coupling \beq {\cal
L}_{u}=-2y_1\left((V_{CKM})^{i3}~\frac{S-iA}{\sqrt{2}}~
\bar{u}_L^{i} u_R^1-H^- ~\bar{b}_L u_R^1 \right)+h.c.\,.
 \label{yukawau}
\eeq

In the quasi-inert Higgs doublet model, both $S$ and $A$
couple to the up quark, which leads that they are no longer the candidate for dark matter.
To provide a candidate for dark matter, a possible approach is to introduce a single scalar field
with an extra discrete symmetry being imposed, which will be studied
elsewhere.

\section{theoretical and experimental constraints}

$\bullet${\bf$D-\bar{D}$ mixing and same-sign top pair production at
LHC:} The couplings of $S$ and $A$ to up quark in Eq.
(\ref{yukawau}) can contribute to the $D-\bar{D}$ mixing and the
same-sign top pair production at LHC \cite{QIHDM,11062142}. The
contributions of $S$ and $A$ are destructive and such contributions
can be canceled for degenerate $S$ and $A$ masses.
Therefore, in this paper, we take $m_S=m_A$, which implies
$\lambda_5=0$ from the Eq. (\ref{lambds}).

$\bullet${\bf Electroweak precision tests:} The electroweak S and T
parameters can give the constraints on the splitting of the charged
Higgs and the neutral Higgs masses \cite{11074350,QIHDM}.
Considering the LEP constraints \cite{lep}, we take
$m_{H^\pm}=m_S=m_A\geq 90$ GeV, which implies $\lambda_4=0$ from the
Eq. (\ref{lambds}).

$\bullet${\bf Vacuum stability and unitarity:} For
$m_{H^\pm}=m_S=m_A$, the vacuum stability can give the constraints
on the parameters \cite{stability}:
\begin{equation}
\lambda_{1,2}>0\,,\qquad\lambda_{3} > - 2\sqrt{\lambda_1
\lambda_2}\,.\label{eq:vacuumstability}
\end{equation}

The unitarity constraints for the scalar potential were calculated in Ref. \cite{12012644,unitarity}.
Various processes were used to constrain quartic couplings at the tree level. These lead to a set
of unitarity constrained parameters:
\begin{eqnarray}
&&e_{1,2}=\lambda_3 \pm \lambda_4 \quad , \quad
e_{3,4}= \lambda_3 \pm \lambda_5 \nonumber\\
&&e_{5,6}= \lambda_3+ 2 \lambda_4 \pm 3\lambda_5\quad , \quad
e_{7,8}=-\lambda_1 - \lambda_2 \pm \sqrt{(\lambda_1 - \lambda_2)^2 + \lambda_4^2}
\nonumber\\
&&
e_{9,10}= -3\lambda_1 - 3\lambda_2 \pm \sqrt{9(\lambda_1 - \lambda_2)^2 + (2\lambda_3 +
   \lambda_4)^2}
\nonumber\\
&&
e_{11,12}=
 -\lambda_1 - \lambda_2 \pm \sqrt{(\lambda_1 - \lambda_2)^2 + \lambda_5^2}.\nonumber
\end{eqnarray}
The absolute values of the twelve parameters must be
smaller than $8\pi$. The strongest constraint on $\lambda_{1,2}$ comes from $e_{9,10}$,
which gives approximately \cite{12012644}:
\beq
\lambda_1 + \lambda_2 < \frac{8 \pi}{3}.
\eeq
The coupling $\lambda_1$ can be determined
by the Higgs mass $m_h$ from the Eq. (\ref{lambds}). Requiring
the absolute values of $e_i$ $(i=1,\cdot\cdot\cdot,12)$ to be smaller than $8\pi$ and
the Eq. (\ref{eq:vacuumstability}) to be satisfied, we can get
the lower bound of $\lambda_3$ for a fixed $m_h$ and $m_{H^\pm}=m_S=m_A$.
For $m_h=120$ GeV, 125 GeV and 130 GeV ($i.~ e.
~\lambda_1=0.12,~0.13,~0.14$), the lower bound of $\lambda_3$ is respectively -1.4,
-1.45 and -1.5.

$\bullet${\bf The $t\bar{t}$ total production cross sections and
$t\bar{t}$ invariant mass distribution at the Tevatron:} The current
cross section measured at Tevatron is $\sigma^{exp}=7.50\pm0.48$ pb
for $m_t=172.5$ GeV \cite{11070841-46}, while the SM cross section
is $\sigma^{SM}=7.46^{+0.66}_{-0.80}$ pb from \cite{11070841-47} and
$\sigma^{SM}=6.30\pm0.19^{+0.31}_{-0.23}$ pb from
\cite{11070841-48}. Here, we conservatively require
$-0.12<\frac{\sigma^{NP}}{\sigma^{SM}}<0.3$ for the Tevatron. The
invariant mass distribution was also measured by CDF, and the
results are presented in nine bins of $M_{t\bar{t}}$
\cite{09121447-22}. We require the differential cross section in
each bin to be within the $2\sigma$ regions of their experimental
values. In this paper, the mass of top quark is taken as 172.5 GeV.

$\bullet${\bf New top quark decay channels:} The top quark can decay into
a light quark and a scalar particle for the scalar mass is light
enough. The measurement of the total top quark width is
$\Gamma_t^{exp}=1.99^{+0.69}_{-0.55}$ GeV \cite{11074350-49}, which is
in agreement with the SM value $\Gamma_t^{SM}=1.3$ GeV. We require
the total top quark width to be within the $2\sigma$ regions of the
experimental value.

The flavor changing couplings in the Eq. (\ref{yukawau}) can
contribute to the single top quark production through the $gu\to tS~(A)$
process. The D0 has recently measured single top quark production cross
section at the Tevatron and obtained $\sigma(p\bar{p}\to tqb+X)=2.90\pm0.59$
pb \cite{11052788}. However, the single top quark production data gives no
constraints to this model because of the difference between final states
in experiments and those in this model.

\section{diphoton rate for a $120-130$ GeV Higgs at the LHC}

In our calculations, we take $m_{H^\pm}=m_S=m_A=m_2\geq 90$ GeV, so
the decays into these Higgs bosons are kinematically forbidden for
a light SM-like Higgs boson $h$. Except for the decay $h\to
\gamma\gamma$, the decay modes and their widths of the SM-like Higgs
boson are the same both in QIHDM and SM. Since the decay width of
$h\to \gamma\gamma$ is very small, the total width of SM-like Higgs
boson in QIHDM almost equals to that in SM.

In the SM, the decay $h\to \gamma\gamma$ is dominated by the W loop
which can also interfere destructively with the subdominant top
quark loop. In QIHDM, the charged Higgs boson $H^{\pm}$ loop can
give the additional contributions to the the decay width
$\Gamma(h\to \gamma\gamma)$, which can be expressed as
\cite{hrr1loop}
\begin{eqnarray}
\Gamma(h\to \gamma\gamma) = \frac{\alpha^2
  m^3_{h}}{256\pi^3v^2}\Bigg| \sum_{i} N_{ci} Q^2_{i} F_{1/2}(\tau_f)
+ F_{1}(\tau_W)+ g_{_{H^{\pm}}}F_{0}(\tau_{H^\pm}) \Bigg|^2 ,
\label{widths}
\end{eqnarray}
where \beq g_{_{H^{\pm}}}=\frac{\lambda_3
v^2}{2m_{H^\pm}^2}=\frac{m_{H^\pm}^2-\mu_2^2}{m_{H^\pm}^2},~~
\tau_f=\frac{4m_f^2}{m_h^2},~~\tau_W=\frac{4m_W^2}{m_h^2},~~
\tau_{H^\pm}=\frac{4m_{H^\pm}^2}{m_h^2}.\eeq $N_{ci}$, $Q_i$ are the
color factor and the electric charge respectively for fermion $i$
running in the loop. The dimensionless loop factors for particles of
spin given in the subscript are:
\begin{eqnarray}
F_1 = 2+3\tau + 3\tau(2-\tau)f(\tau), \quad F_{1/2} =
-2\tau[1+(1-\tau)f(\tau)], \quad F_0 = \tau[1-\tau f(\tau)],
\end{eqnarray}
with
\begin{equation}
f(\tau) = \left\{ \begin{array}{lr}
[\sin^{-1}(1/\sqrt{\tau})]^2, & \tau \geq 1 \\
-\frac{1}{4} [\ln(\eta_+/\eta_-) - i \pi]^2, & \, \tau < 1
\end{array}  \right.
\end{equation}
where $\eta_{\pm} = 1 \pm \sqrt{1-\tau}$.

\begin{figure}[tb]
 \epsfig{file=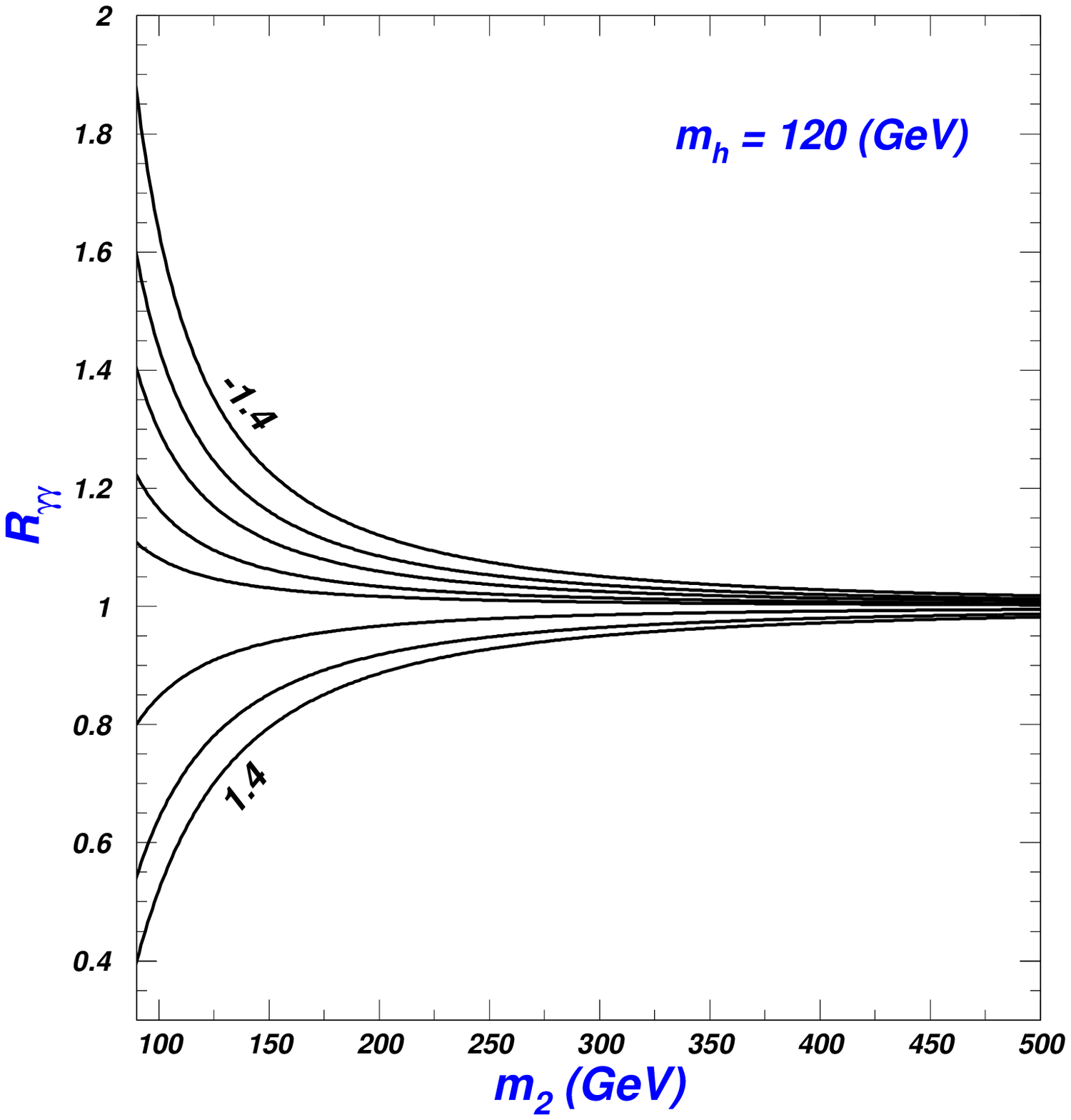,height=5.8cm}
 \epsfig{file=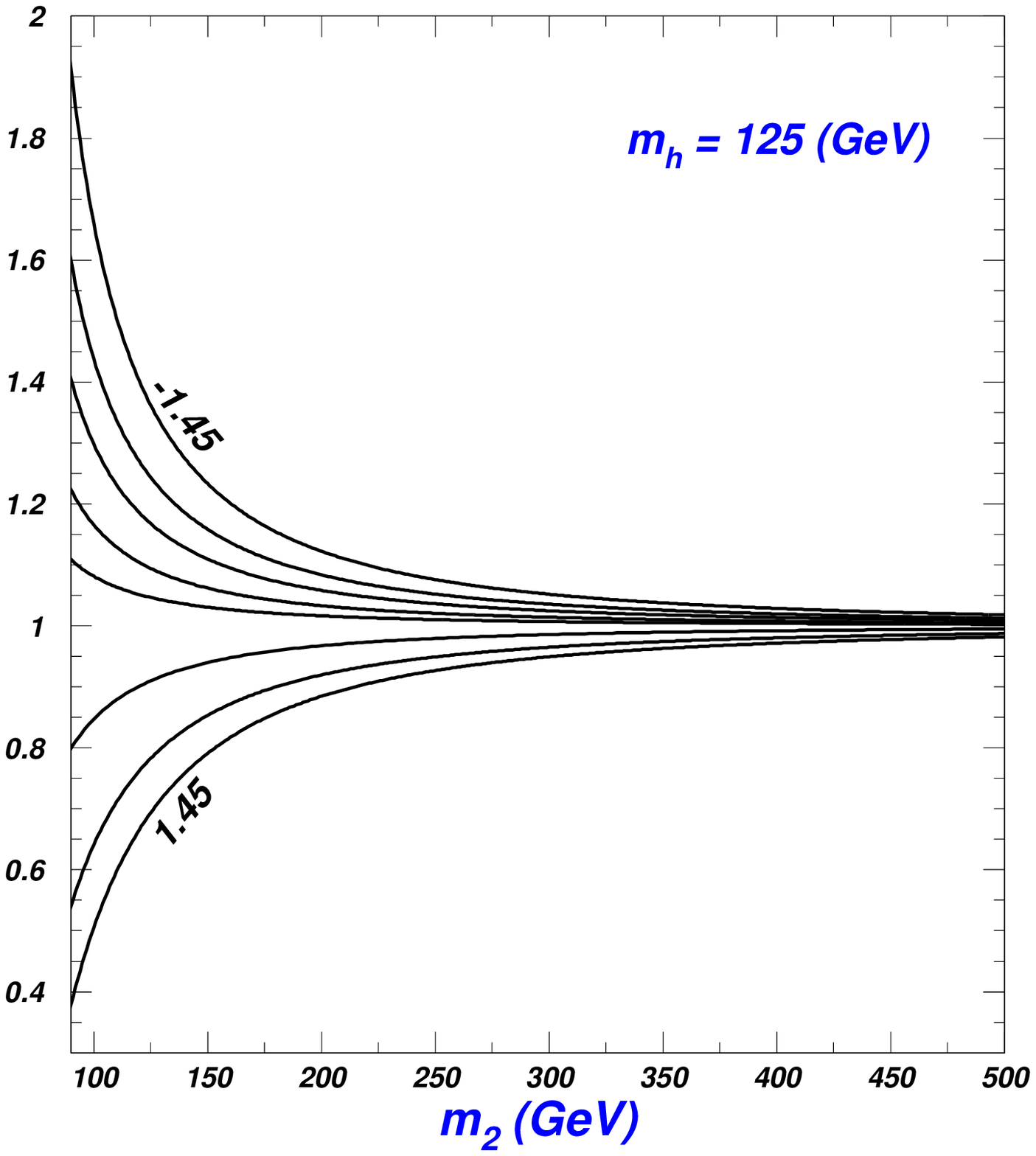,height=5.8cm}
 \epsfig{file=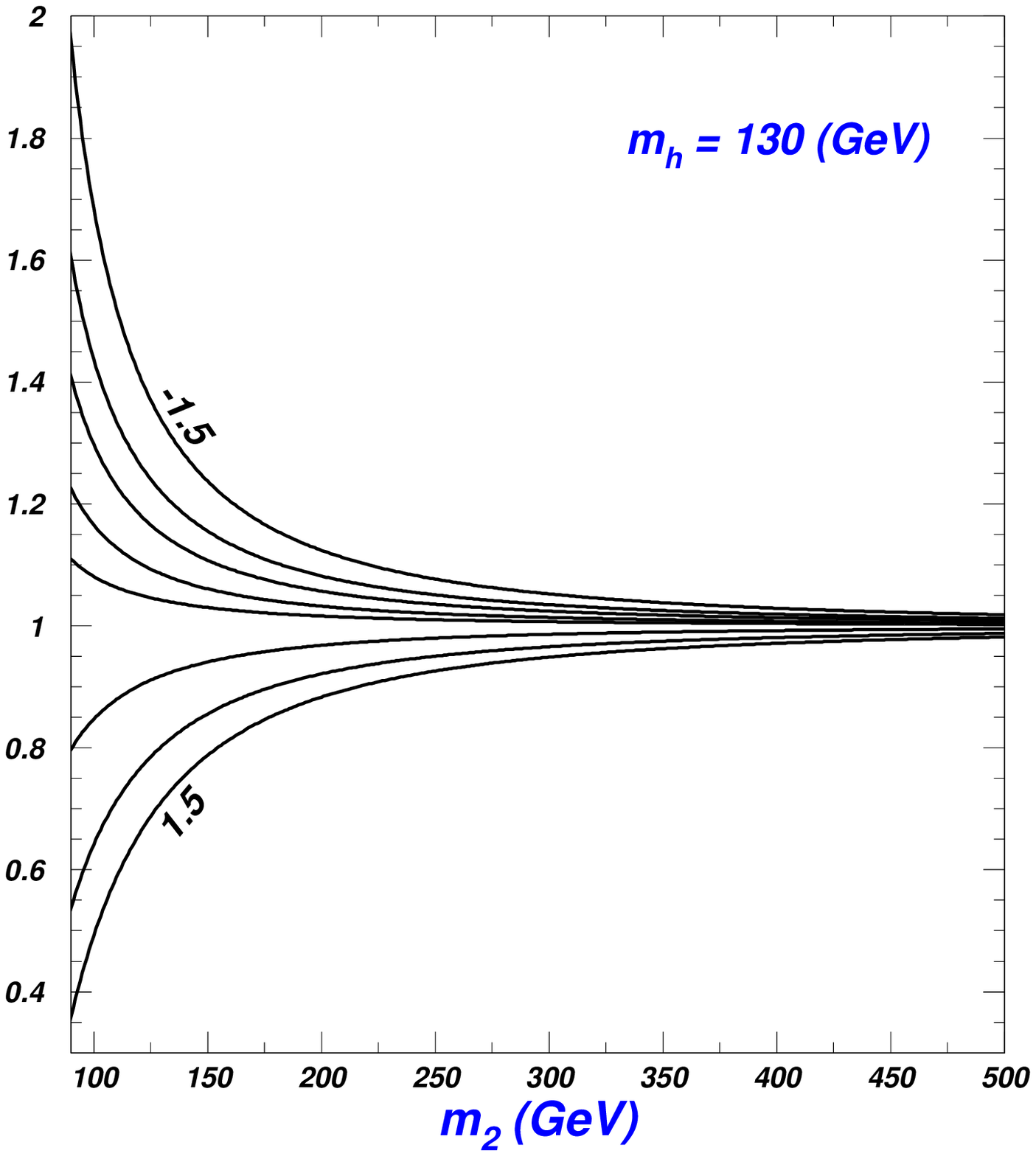,height=5.8cm}
\vspace{-0.2cm} \caption{The $R_{\gamma\gamma}$ versus the $H^\pm$
mass ($m_2$) for several different $\lambda_3$. For the left
(middle, right) panel, $m_h=120$ (125, 130) GeV and the curves from
bottom to top correspond to $\lambda_3=$1.4 (1.45,
1.5),~1.0,~0.4,~-0.2,~-0.4,~-0.7,~-1.0,~-1.4 (-1.45, 1.5).}
\label{hrr}
\end{figure}

The Higgs boson production cross section at the LHC are the same
both in the QIHDM and SM. Therefore, the LHC diphoton rate of Higgs
boson in the QIHDM normalized to the SM prediction can be written as
\beq
R_{\gamma\gamma}=\frac{Br(h\to\gamma\gamma)}{Br(h\to\gamma\gamma)^{SM}}
\simeq\frac{\Gamma(h\to\gamma\gamma)}{\Gamma(h\to\gamma\gamma)^{SM}}.\eeq

In Fig. \ref{hrr} we show the $R_{\gamma\gamma}$ versus the $H^\pm$
mass ($m_2$) for several different $\lambda_3$. Fig. \ref{hrr} shows
$R_{\gamma\gamma}$ is almost the same for $m_h=$120 GeV, 125 GeV and
130 GeV. $H^{\pm}$ contributions can interfere constructively for
$\lambda_3<0$ and interfere destructively for $\lambda_3>0$, leading
$R_{\gamma\gamma}>1$ and $R_{\gamma\gamma}<1$, respectively. The
magnitude becomes sizable as the decreasing of the $H^\pm$ mass
since $g_{_{H^{\pm}}}$ is proportional to the $1/m_{H^\pm}^2$. For
$m_2 =90$ GeV and $\lambda_3$ in the range of -0.2 and -1.5,
$R_{\gamma\gamma}$ varies from 1.1 to 1.95.

\section{top quark forward-backward asymmetry at Tevatron}

\begin{figure}[tb]
 \epsfig{file=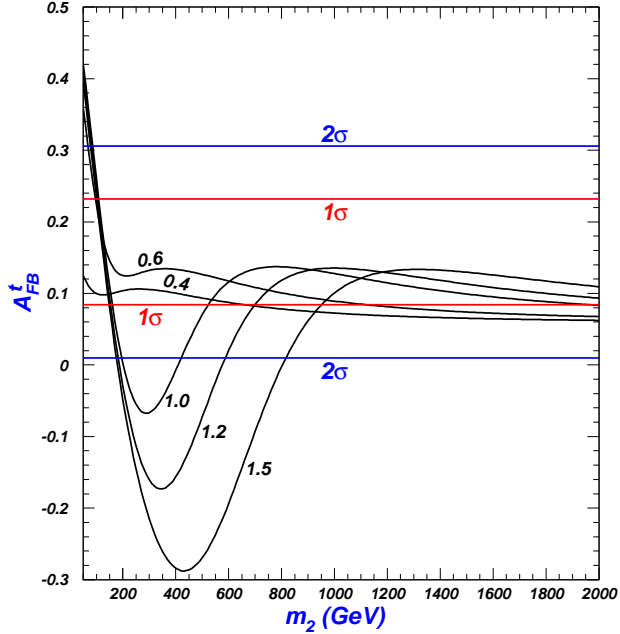,height=8.5cm}
\vspace{-.5cm} \caption{The top forward-backward asymmetry
$A_{FB}^t$ at Tevatron versus $m_{2}$ for several different $y_1$.
The horizontal lines show the $1\sigma$ and $2\sigma$ ranges from
the corresponding experimental data.} \label{afbys}
\end{figure}

In the $t\bar{t}$ rest frame at Tevatron, the top quark
forward-backward asymmetry $A_{FB}^t$ is defined by \cite{afbdefi}
\beq A_{FB}^t=A_{FB}^{NP}\times R + A_{FB}^{SM}\times (1-R)
\label{afbgongsi} \eeq where $A_{FB}^{SM}=0.058$ is the asymmetry in
the SM, and
\begin{eqnarray}
&& A_{FB}^{NP}=\frac{\sigma^{NP}(\Delta y >0)
-\sigma^{NP}(\Delta y <0)}{\sigma^{NP}(\Delta y >0)+\sigma^{NP}(\Delta y <0)},\\
&& R=\frac{\sigma^{NP}}{\sigma^{SM}+\sigma^{NP}} \eea
are the
asymmetry induced by the new physics and the fraction of the new
physics contribution to the total cross section, respectively.
$\Delta y$ is the rapidity difference between a top and an anti-top.
In our calculations, we take $m_t=172.5$ GeV and use the parton
distribution function CTEQ6L \cite{cteq} with renormalization scale
and factorization scale $\mu_R = \mu_F=m_t$. We assume that the
K-factors are universal, so that the QCD correction effects are
canceled in the ratios of $\sigma^{NP}/\sigma^{SM}$ and
$\sigma^{NP}/(\sigma^{SM}+\sigma^{NP})$, and they are the same at LO
and NLO.

The matrix elements $M$ of the process $u(p_1)\bar{u}(p_2)\to
t(k_1)\bar{t}(k_2)$, including the SM, new scalar $S$ and $A$
contributions, can be written as ref. \cite{09113237} \beq
\sum|M|^2=\frac{16g_s^4}{s^2}(t_t^2+u_t^2+2sm_t^2)+32g_s^2y^2\frac{sm_t^2+t_t^2}{st_{m_2}}
+36\frac{y^4t_t^2}{t^2_{m_2}}, \label{amph0}\eeq where
$s=(p_1+p_2)^2$, $t=(p_1-k_1)^2$, $u=(p_1-k_2)^2$, $t_t=t-m_t^2$,
 $t_{m_2}=t-m^2_{2}$, $y=\sqrt{2}y_1$.

\begin{figure}[tb]
 \epsfig{file=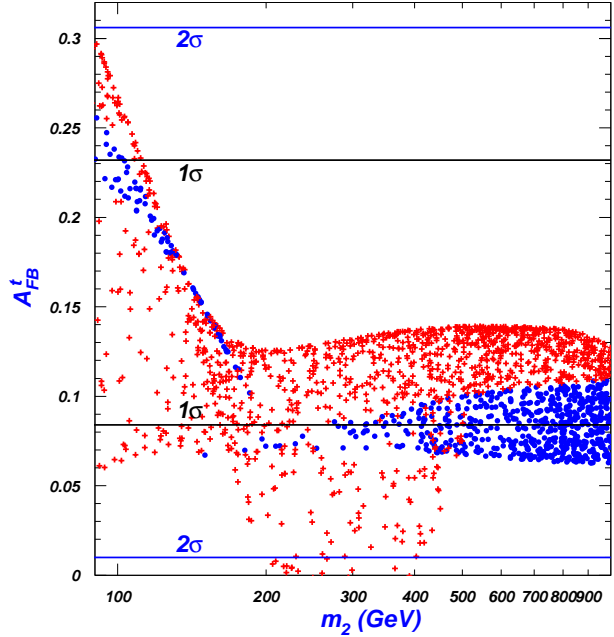,height=8.5cm}
\vspace{-0.3cm} \caption{The top quark forward-backward asymmetry
$A_{FB}^t$ at Tevatron versus $m_2$. The bullets (blue) and crosses
(red) are respectively allowed and excluded by the three
experimental data of top quark. The horizontal lines show the
$1\sigma$ and $2\sigma$ ranges from the corresponding experimental
data.} \label{afbyn}
\end{figure}

\begin{figure}[tb]
 \epsfig{file=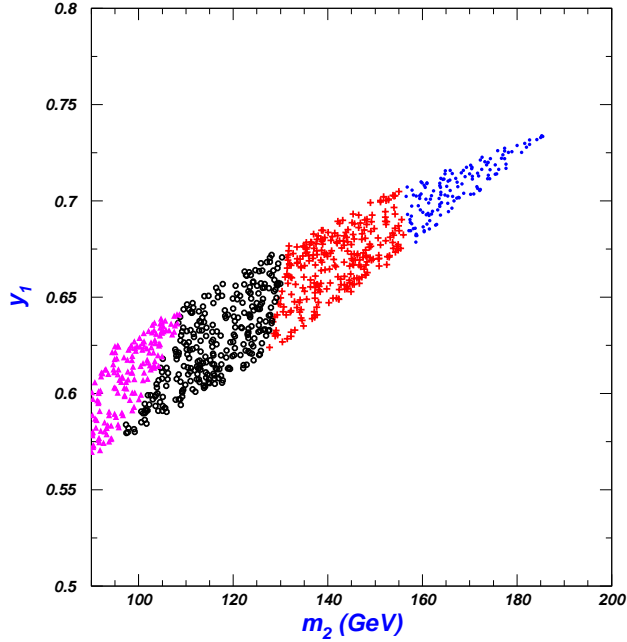,height=8.5cm}
\vspace{-0.3cm} \caption{Scatter plots for $y_1$ versus $m_2$, which
are allowed by the experimental data of top quark. $A_{FB}^t$ is in
the range of $0.1-0.14$ for bullets (blue), $0.14-0.18$ for crosses
(red), $0.18-0.22$ for circles (black) and $0.22-0.25$ for triangle
(pink), respectively.} \label{ymh}
\end{figure}

In Fig. \ref{afbys}, we plot the top quark forward-backward
asymmetry $A_{FB}^t$ at Tevatron versus the masses of $S$ and $A$
($m_2$) for several different $y_1$. We see that $A_{FB}^t$ can be
enhanced sizably for the very low value of $m_2$, be over 0.1 for
the large one and be negative in the intermediate region.

In  Fig. \ref{afbyn}, we scan the following parameter space, \beq
0.2 \leq y_1 \leq 1.0,~~~90~ \rm GeV \leq m_{2} \leq 1000~\rm GeV,
\label{scan}\eeq and plot $A_{FB}^t$ versus $m_2$ taking into
account the total decay width of top quark, top quark pair
production cross section and $t\bar{t}$ mass distribution at
Tevatron. We find that the relative small parameter space scanned is
allowed by the above experimental data of top quark, where
$A_{FB}^t$ can be explained to within $1\sigma$ and reach 0.25. Our
numerical results show that the $t\bar{t}$ cross section and
invariant mass distribution measured at Tevatron give the most
constraints on the parameters $y_1$ and $m_2$. The total top quark
width can hardly give further constraints.

For the large $m_2$, Fig. \ref{afbyn} shows $A_{FB}^t$ is below
0.12,  and Fig. \ref{hrr} shows $R_{\gamma\gamma}$ is close to 1.
For $m_2<$ 200 GeV, both $A_{FB}^t$ and $R_{\gamma\gamma}$ can be
enhanced sizably. Therefore, we scan \beq 0.45 \leq y_1 \leq 1.0,
~~~90~ \rm GeV \leq m_2\leq 200~ \rm GeV , \eeq and give Fig.
\ref{ymh} and the right panel of Fig. \ref{hrrafb}. We exclude the
parameters which are not in agreement with the experimental data of
top quark. Fig. \ref{ymh} shows the $y_1$ and $m_2$ for which
$A_{FB}^t$ can reach $0.1-0.25$. We can see that $0.58\leq
y_1\leq0.7$ and 100 GeV $\leq m_2\leq 155$ GeV are favored for $0.14
<A_{FB}^t< 0.22$. A very small $y_1$ is allowed by the experimental
data of top quark, but it will lead $A_{FB}^t$ to be much less than
0.1, which are not shown in Fig. \ref{ymh}.

The right frame of Fig. \ref{hrrafb} shows $m_2$ versus $A_{FB}^t$.
To examine the correlation between $A_{FB}^t$ and the diphoton rate
of the SM-like Higgs boson at LHC, we plot $m_2$ versus
$R_{\gamma\gamma}$ for $m_h=125$ GeV in the left panel of Fig.
\ref{hrrafb}. The correlation between $A_{FB}^t$ and
$R_{\gamma\gamma}$ depends on $m_2$, which can contribute to the two
observables. For example, for $A_{FB}^t=0.16$, the right panel of
Fig. \ref{hrrafb} shows that $m_2$ should be around 140 GeV, while,
for such value of $m_2$, the left panel of Fig. \ref{hrrafb} shows
that $R_{\gamma\gamma}$ is less than 1.3. By the same way,
$R_{\gamma\gamma}$ should be less than 1.2 for $A_{FB}^t=0.12$, and
$A_{FB}^t$ can reach $0.21-0.26$ for $R_{\gamma\gamma}=1.6$. The
larger $A_{FB}^t$ implies that the diphoton rate can be enhanced
more sizably with respect to the SM prediction. Moreover, the larger
$R_{\gamma\gamma}$ can imply more precisely the range of $A_{FB}^t$.

\begin{figure}[tb]
 \epsfig{file=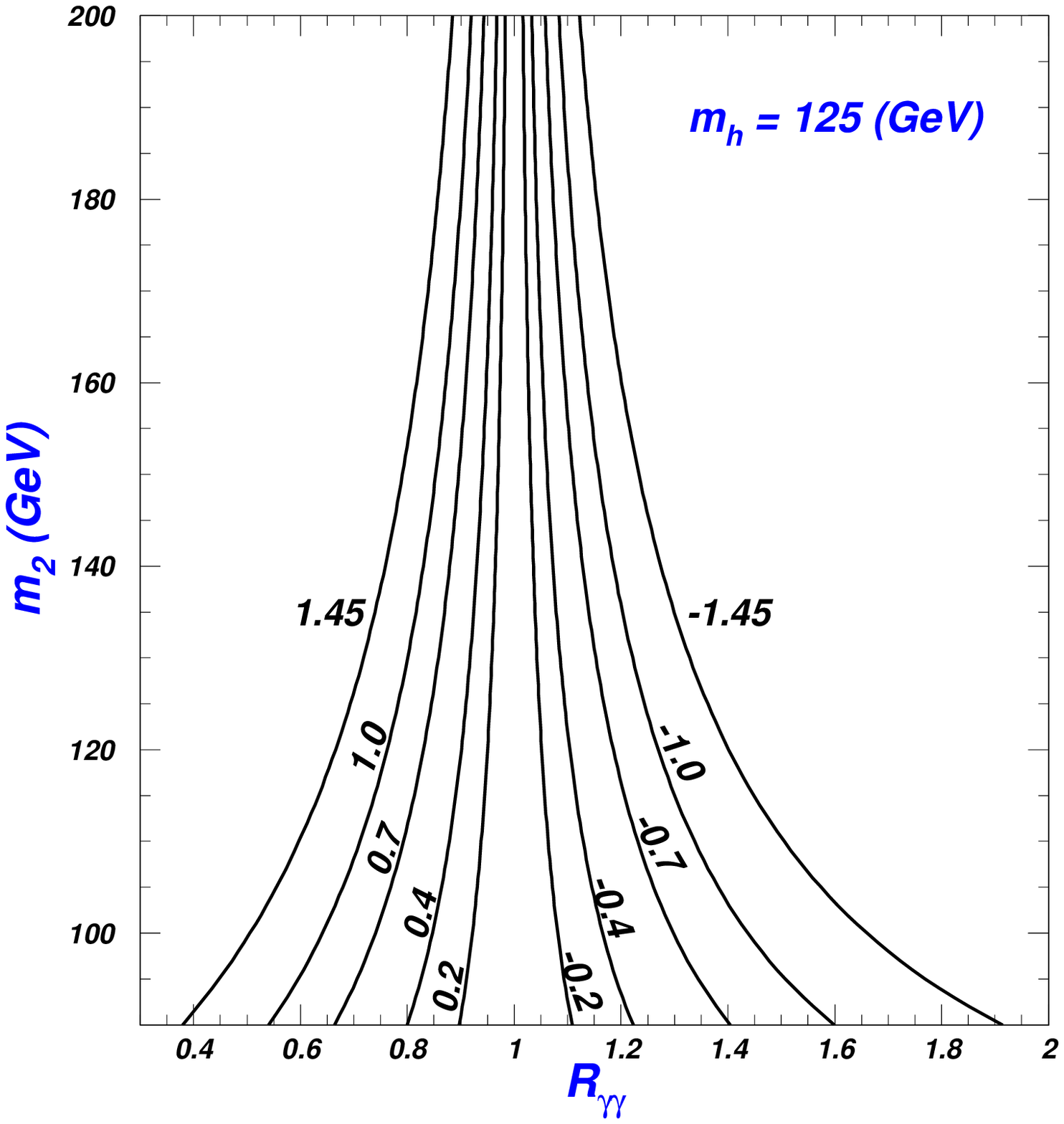,height=7.0cm}
  \epsfig{file=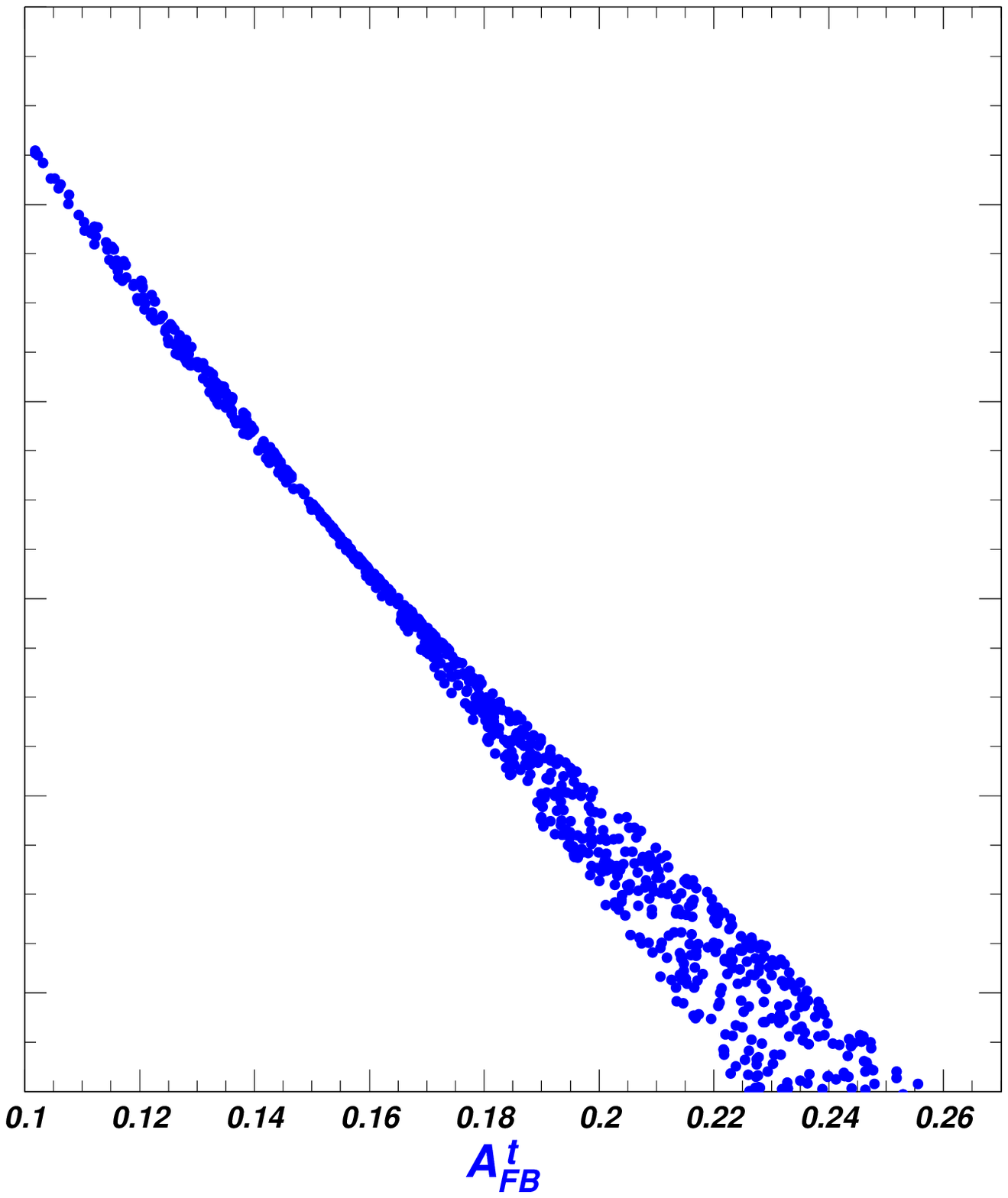,height=7.0cm}
\vspace{-0.3cm} \caption{The left panel: $m_2$ versus
$R_{\gamma\gamma}$ for $m_h=125$ GeV. The right panel: Scatter plots
for $m_2$ versus $A_{FB}^t$.} \label{hrrafb}
\end{figure}

Note that the correlation between $A_{FB}^t$ and
$R_{\gamma\gamma}$ found in the above discussions
is due to the lower bound on $\lambda_3$ coming from the
requirement of vacuum stability and unitarity, and from the assumption
that the scalars $S$, $A$ and $H^\pm$ are degenerate in mass.
Ref. \cite{QIHDM} shows that, for a few GeV splitting between $m_S$ and $m_A$,
the present experimental data of $D-\bar{D}$ mixing requires
the coupling constant $y_{uc}=2y_1(V_{CKM})^{23}$ to be smaller than $10^{-3}$,
 i.e. $y_1 < 10^{-2}$. Therefore, the scalars $S$ and $A$ should be degenerate
in mass so that $y_1$ is allowed to be enough large to explain $A_{FB}^t$.
However, the electroweak S and T parameters allow the mass of the charged scalar to be shifted
from the masses of the neutral scalars by at most $\sim$ 110 GeV \cite{11074350}.
For the light scalars, the correlation between $A_{FB}^t$ and $R_{\gamma\gamma}$
will be greatly softened and even disappear due to such mass splitting.
For example, for the masses of the neutral scalars are in the range 90 $-$ 200 GeV,
the mass of the charged scalar can be as low as 90 GeV, i.e.
for 0.1 $ < A_{FB}^t <$ 0.26, the maximal value of $R_{\gamma\gamma}$ is always
1.95 (where the mass splitting gives a negligible effect on the lower bound of $\lambda_3$).
Fig. \ref{hrrafb} shows that the charged scalar and neutral scalars with less than 150 GeV mass
can respectively fit the experimental data of $R_{\gamma\gamma}$ and $A_{FB}^t$ well.
Therefore, the small mass splitting between the charged scalar and the neutral scalars is
favored by the experimental data.

\section{Conclusion}
In the framework of quasi-inert Higgs doublet model, we study the
LHC diphoton rate for a SM-like Higgs boson and the top quark
forward-backward asymmetry $A_{FB}^t$ at Tevatron. Taking the
theoretical and experimental constraints, we find that the diphoton
rate of Higgs boson at LHC can be enhanced sizably due to the light
charged Higgs contributions with respect to the SM prediction, while
the measurement of the top quark forward-backward asymmetry at
Tevatron can be explained to within $1\sigma$ due to the non-SM
neutral Higgs bosons contributions. Besides, there are some
correlations between the two observables. Compared to the SM
prediction, the diphoton rate can be enhanced more sizably for the
larger top quark forward-backward asymmetry. Furthermore, the range
of $A_{FB}^t$ can be narrowed more sizably for the larger
$R_{\gamma\gamma}$.

\section*{Acknowledgment}
This work was supported by the National Natural Science Foundation
of China (NNSFC) under grant Nos. 11105116 and 11005089.

\end{document}